\documentclass[12pt]{article}
\usepackage{latexsym}
\usepackage{amsmath}

\author{\large\bf Metin Ar\i{}k and Mikhail B. Sheftel\\
Department of Physics, Bo\u{g}azi\c{c}i University\\ 34342 Bebek,
Istanbul, Turkey}
\title{\Large \bf Symmetry analysis and exact solutions of modified Brans-Dicke
cosmological equations}

\date{}
\begin{document}
\maketitle

\begin{abstract}
 We perform a symmetry analysis of modified Brans-Dicke
 cosmological equations and present exact solutions. We discuss
 how the solutions may help to build models of cosmology where,
 for the early universe, the expansion is linear and the equation
 of state just changes the expansion velocity but not the linearity.
 For the late universe the expansion is exponential and the
 effect of the equation of state on the rate of expansion is just
 to change the constant Hubble parameter.
\end{abstract}

\section{Introduction}

The standard model of cosmology \cite{roos} has undergone several
modifications in the past twenty years. With the discovery
\cite{PenWils} of the cosmic microwave background radiation the
early cosmological models \cite{Fried,Lem} developed into the
first standard model which had the early radiation dominated and
the late dust dominated stages. Observational and theoretical
consistency has forced modifications on the standard model.  We
now believe that the universe is born in a radiation dominated
stage, then undergoes an inflationary exponentially expanding
stage, then becomes radiation dominated again, and then continues
with matter domination, which by today has evolved into a dark
energy dominated, slowly but exponentially expanding, stage. The
main reason for such a complicated history of domination is that
according to Einstein cosmological field equations the rate of
expansion of the universe depends on the equation of state of the
matter-energy that fills it. One immediate question which arises
is whether there is any consistent modification of Einstein's
equations such that the expansion of the universe is independent
of its content. In this paper we would like to study such a model.
The model consists of a modified Brans-Dicke-Jordan-Thirry
\cite{JTBD} model where the signs of the kinetic and potential
terms of the "scalar field" are negative when written to the right
hand (wooden) side of the Einstein equations but is positive just
as the geometric term containing the square of the time derivative
of the scale size of the universe when written to the left hand
(marble side). Thus the Friedman equation
\begin{equation}\label{1}
 3\left(\frac{\dot{a}^2}{a^2} + \frac{k}{a^2}\right) = 8\pi G\rho
\end{equation}
is changed into
\begin{equation}\label{2}
 3\left(\frac{\dot{a}^2}{a^2} + \frac{k}{a^2}\right) + 2\omega\frac{\dot{\phi}^2}{\phi^2} + \dots =
 \frac{4\omega}{\phi^2}\rho
\end{equation}
where $\phi$ is a new geometric field inspired by the
Brans-Dicke-Jordan-Thirry model. The reason we call $\phi$ a
geometric field is that its appearance is quite similar to the
scale size $a$ of the universe. The terms $\dots$ will be
determined by the covariant action. In fact, the analogous
equation for a Kaluza-Klein theory with $n$ internal dimensions is
\cite{DT}
\begin{equation}\label{3}
 3\left(\frac{\dot{a}^2}{a^2} + \frac{k}{a^2}\right) + \frac{1}{2}\, n(n-1)\frac{\dot{b}^2}{b^2}+ \dots = 8\pi G\rho
\end{equation}
where $b$ is the scale size of the internal manifold.

In section \ref{sec:eqs} we present the basic equations of the
model as applied to cosmology. In section \ref{sec:symmetry} we
perform a  symmetry analysis based on dilatational symmetries and
show that symmetries exist under the conditions that the mass of
the "scalar field" vanishes and/or the curvature of spacelike
sections is zero. The solutions are analyzed in sections
\ref{sec:k=0}, \ref{sec:m=0} and \ref{sec:k=m=0}.

In section \ref{conclude} we conclude with a model of expansion
for the universe where, in the early stages, the mass term of the
"scalar field" can be neglected $(m=0)$ but the curvature term is
important. For closed $(k=1)$ spacelike sections the universe
expands linearly. This linear expansion is independent of the
equation of state. However, the rate of change of the newtonian
gravitational constant depends on the equation of state. This is
in a strong contrast to standard general relativistic cosmology
where the rate of expansion strongly depends on the equation of
state.

Whereas the curvature term is important in the early stages, we
argue that in later stages, after the universe has expanded to a
large size, it can be neglected. The mass term then causes a slow
exponential expansion which can be identified with the dark energy
phenomenon of the present day universe. This phenomenon is also
independent of the equation of state which only slightly
influences the exponential rate of expansion.

\section{Basic equations of the model}
\setcounter{equation}{0}
 \label{sec:eqs}

The action is the following:
\begin{equation}\label{action}
 S = {\displaystyle\int} {\rm d}^4 x\,\sqrt{g}\left[
 -\frac{1}{8\omega}\phi^{2}R - \frac{1}{2}g^{\mu\nu}
 \partial_{\mu}\phi\partial_\nu \phi + \frac{1}{2}\,m^2\phi^2
 + L_M\right],
\end{equation}
where $\phi$ represents the Brans-Dicke scalar field, $\omega$
denotes the dimensionless Brans-Dicke parameter taken to be much
larger than $1$, $\omega\gg1$. The scalar field has a kinetic term
$- \frac{1}{2}g^{\mu\nu}
 \partial_{\mu}\phi\partial_\nu \phi$ with the wrong sign and the potential
 of the scalar field
contains only a mass term with the wrong sign
$\frac{1}{2}m^{2}\phi^{2}$. $L_M$, on the other hand, is the
matter Lagrangian such that the scalar field $\phi$ does not
couple with it. $R$ is the Ricci scalar. For simplicity we also
restrict our analysis to the Robertson Walker metric to emphasize
that $\phi$ is necessarily spatially homogeneous:
\begin{equation}\label{metric}
{\rm d}s^2 = \rm{ d}t^2 -a^2(t)\frac{{\rm
d}\vec{x}^{\;2}}{\left[\displaystyle 1 +
\frac{k}{4}\vec{x}^{\;2}\right]^2},
\end{equation}
where $k$\ is the curvature parameter with $k=-1$, $0$, $1$
corresponding to open, flat, closed universes respectively and
$a\left( t\right)  $ is the scale factor of the universe. After
applying the variational procedure to the action and assuming
$\phi=\phi\left(  t\right)  $ and energy momentum tensor of matter
and radiation excluding $\phi$ is in the perfect fluid form of
$T_{\nu}^{\mu}=diag\left(  \rho,-p,-p,-p\right)  $ where $\rho$ is
the energy density and $p$ is the pressure and also noting that
the right hand side of the $\phi$ equation must be zero in
accordance with our previous argument on $L_{M}$ being independent
of $\phi$, the field equations reduce to (dots denote
$\rm{d}/\rm{d}t$)
\begin{equation}\label{dens}
 \frac{3}{4\omega}\phi^2\left(\frac{\dot{a}^2}{a^2} + \frac{k}{a^2}\right)
 + \frac{1}{2}\dot{\phi}^2 + \frac{1}{2}m^2\phi^2 + \frac{3}{2\omega}
 \frac{\dot{a}}{a}\dot{\phi}\phi = \rho
\end{equation}
\begin{equation}\label{pres}
 \frac{-1}{4\omega}\phi^2\left(2\frac{\ddot{a}}{a} +
 \frac{\dot{a}^2}{a^2} + \frac{k}{a^2}\right) -
 \frac{1}{\omega}\frac{\dot{a}}{a}\frac{\dot{\phi}}{\phi} -
 \frac{1}{2\omega}\ddot{\phi}\phi + \left(\frac{1}{2} - \frac{1}{2\omega}
 \right)\dot{\phi}^2 - \frac{1}{2}m^2\phi^2 = p
 \end{equation}
\begin{equation}\label{fi}
 \ddot{\phi} + 3\frac{\dot{a}}{a}\phi + \left[m^2 + \frac{3}{2\omega}
 \left(\frac{\ddot{a}}{a} + \frac{\dot{a}^2}{a^2} + \frac{k}{a^2}\right)
 \right]\phi = 0 .
\end{equation}
The following identity shows that the continuity equation could be
considered as a consequence of these three equations:
\begin{equation}\label{cont}
  -\left[\frac{\rm{d}}{\rm{d}t}(\ref{dens}) + 3\frac{\dot{a}}{a}\times\left((\ref{dens})
  + (\ref{pres})\right) - \dot{\phi}\times(\ref{fi})\right] \equiv
  \dot{\rho} + 3\frac{\dot{a}}{a}\,(p + \rho) = 0.
\end{equation}
Instead, we prefer to consider more complicated pressure equation
(\ref{pres}) as an algebraic consequence of
${\displaystyle\frac{\rm{d}(\ref{dens})}{\rm{d}t}}$, (\ref{fi}),
(\ref{dens}) and the continuity equation in (\ref{cont}).

We assume the power law for the density
\begin{equation}\label{power}
  \rho = C a^\nu \quad \Longrightarrow \quad \dot{\rho} = \nu\rho
  H
\end{equation}
which implies that the continuity equation (\ref{cont}) becomes
the equation of state:
\begin{equation}\label{stateq}
  p = - \frac{(\nu + 3)}{3}\,\rho,
\end{equation}
so that for dust we have $\nu = -3,\, p = 0$ and for radiation
$\nu = -4,\, p = \rho/3$.

Now define new unknowns $F(a)$ and $H(a)$ by the formulas
\begin{equation}\label{def}
  F = \frac{\dot{\phi}}{\phi}\,, \quad H = \frac{\dot{a}}{a}
\quad \Longrightarrow \quad \frac{\ddot{a}}{a} = aHH^\prime +
H^2,\quad \frac{\ddot{\phi}}{\phi} = aHF^\prime + F^2 ,
\end{equation}
where prime denotes derivative of a function of $a$ and the
relations $\dot{H} = H^\prime(a)\dot{a} = aHH^\prime$, $\dot{F} =
F^\prime(a)\dot{a} = aHF^\prime$ were used. Equations
(\ref{dens})--(\ref{fi}) become the equations for unknowns $H(a)$
and $F(a)$
\begin{eqnarray}
 & & H^2 + 2HF + \frac{2\omega}{3}\left(F^2 + m^2\right) +
 \frac{k}{a^2} = \left(\frac{4\omega}{3}\right)\frac{\rho}{\phi^2} \equiv
 \left(\frac{4\omega}{3}\right)\frac{Ca^\nu}{\phi^2}
 \label{HF1}
  \\ & & \frac{2}{3}\,a H\left(H^\prime + F^\prime\right) + H^2 +
  \frac{4}{3}\, H F + \frac{2}{3}(2 - \omega)F^2 + \frac{k}{3a^2}
  + \frac{2\omega}{3}\,m^2 \nonumber \\
   & & = \left(-\frac{4\omega}{3}\right) \frac{p}{\phi^2}
    \label{HF2}
    \\ & & aH\left(\frac{1}{2}\,H^\prime +
  \frac{\omega}{3}\,F^\prime\right) + H^2 + \omega H F +
  \frac{\omega}{3}\,(F^2 + m^2) + \frac{k}{2a^2} = 0 .
  \label{HF3}
\end{eqnarray}
The more lengthy equation (\ref{HF2}) is still an algebraic
consequence of
\,${\displaystyle\frac{\rm{d}(\ref{HF1})}{\rm{d}t}}$, (\ref{HF3}),
(\ref{HF1}) and the continuity equation in (\ref{cont}) due to the
linear relation:
\[ \frac{\rm{d}(\ref{HF1})}{\rm{d}t} + 2F\times (\ref{HF1}) +
3H\times \left((\ref{HF1}) - (\ref{HF2})\right)+ 4F\times
(\ref{HF3}) = 0, \] so we choose (\ref{HF1}) and (\ref{HF3})
together with the state equation (\ref{stateq}) as independent
dynamic equations. However, these equations do not form a closed
system for determining $H(a)$ and $F(a)$ because of $\phi$ in the
right-hand side of (\ref{HF1}). To eliminate $\phi$ from the
$\rho$-equation (\ref{HF1}), we differentiate (\ref{HF1}) with
respect to time, use
\[\dot{a} = aH,\quad \dot{\phi} = F\phi,\quad \dot{H} = aHH^\prime(a),\quad \dot{F} =
aHF^\prime(a) \]
 and eliminate $\phi$ with the aid of (\ref{HF1}) with the result
\begin{eqnarray} \label{1prime}
 & & \left(H^2 + HF\right)H^\prime + \left(H^2 + \frac{2}{3}\,\omega
 HF\right)F^\prime = \frac{\nu}{2a}\,H^3 + \frac{(\nu -
 1)}{a}\,H^2F \\
 & & \mbox{} + \frac{(\nu\omega - 6)}{3a}\,HF^2 -
 \frac{2\omega}{3a}\,F^3 + \frac{k}{a^3}\,\left[\frac{(\nu + 2)}{2}\,H -
 F\right] + \frac{m^2\omega}{3a}\,(\nu H - 2F) \nonumber .
\end{eqnarray}
Solving algebraically equations (\ref{1prime}) and (\ref{HF3})
with respect to $H^\prime$ and $F^\prime$, we obtain the closed
system of two first order equations in normal form that determines
two unknown functions $H(a)$ and $F(a)$:
\begin{eqnarray}
 (2\omega - 3)aH\frac{d H}{d a} &=& (\nu\omega + 6)H^2 +
 2(\nu + 4)\omega H F + \frac{2\omega}{3}\,[(\nu + 6)\omega -
 3]F^2 \nonumber
  \\ & & \mbox{} + \frac{2\omega}{3}\,(\nu\omega + 3)m^2 +
  \frac{k}{a^2}\,[(\nu + 2)\omega + 3] ,
  \label{Heq}
  \\ - (2\omega - 3)aH\frac{d F}{d a} &=& \frac{3}{2}\,(\nu +
  4)H^2 + 3(2\omega + \nu + 1)H F + [(\nu + 8)\omega - 6]F^2
  \nonumber
  \\ & & \mbox{} + (\nu + 2)\omega m^2 + \frac{3k(\nu +
  4)}{2a^2}\, .
  \label{Feq}
\end{eqnarray}
Equation (\ref{HF1}) serves as initial condition for this
dynamical system.

\section{Symmetry group analysis of basic\\ dynamical equations}
\setcounter{equation}{0}
 \label{sec:symmetry}

 We will use classical Lie group analysis \cite{olv} for symmetries of the
system (\ref{Heq}), (\ref{Feq}) as a tool for finding exact
solutions of these equations. Since the equations are of first
order, it is impossible to find all Lie groups of point symmetries
that are admitted by this system (a symmetry condition cannot be
split in the derivatives of unknowns to generate an overdetermined
system, that can be solved and yield all the point symmetries,
because these derivatives are expressed from (\ref{Heq}) and
(\ref{Feq})). Therefore, we make an ansatz for the form of
symmetries that could be admitted by these equations under certain
conditions. We note that equations (\ref{Heq}) and (\ref{Feq}))
are very close to those ones that admit scaling (or dilatational)
symmetries and an obstacle to this is the mass term and/or
curvature term (the one with $k$) in each equation. Indeed, a
scaling symmetry group of transformations of independent and
dependent variables is defined by
\begin{equation}\label{group}
\tilde{a} = \lambda^\alpha a,\quad \tilde{F} = \lambda^\beta F,
\quad \tilde{H} = \lambda^\gamma H .
\end{equation}
It is generated by the infinitesimal generator
\begin{equation}\label{generator}
  X = \alpha a\frac{\partial}{\partial a} + \beta
  F\frac{\partial}{\partial F} + \gamma H\frac{\partial}{\partial
  H}\,.
\end{equation}
Under transformations (\ref{group}), equations (\ref{Heq}) and
(\ref{Feq})) become
\begin{eqnarray}
 & & (2\omega - 3)aH\frac{d H}{d a} = (\nu\omega + 6)H^2 +
 \lambda^{\beta - \gamma} 2(\nu + 4)\omega H F  \nonumber
   \\[1mm] & & \mbox{} +  \lambda^{2(\beta - \gamma)} \frac{2\omega}{3}\,[(\nu + 6)\omega -
 3]F^2 + \lambda^{-2\gamma} \frac{2\omega}{3}\,(\nu\omega + 3)m^2
 \label{Htr}
 \\[1mm] & & \mbox{} + \lambda^{-2(\alpha + \gamma)} \frac{k}{a^2}\,[(\nu + 2)\omega + 3] \nonumber ,
    \\[2mm] & & - (2\omega - 3)aH\frac{d F}{d a} = \lambda^{\gamma - \beta} \frac{3}{2}\,(\nu +
  4)H^2 + 3(2\omega + \nu + 1)H F \nonumber \\[1mm]
  & & \mbox{} + \lambda^{\beta - \gamma} [(\nu + 8)\omega - 6]F^2 + \lambda^{- (\beta + \gamma)} (\nu + 2)\omega m^2
  \nonumber
  \\[1mm] & & \mbox{} + \lambda^{- (2\alpha + \beta + \gamma)} \frac{3k(\nu + 4)}{2a^2}\, .
  \label{Ftr}
\end{eqnarray}
The condition of invariance of equations (\ref{Heq}) and
(\ref{Feq})) under the Lie group of transformations (\ref{group})
is that the transformed equations (\ref{Htr}) and (\ref{Ftr})
should coincide with equations (\ref{Heq}) and (\ref{Feq})), that
is, all the $\lambda$-dependence should vanish.

The first obvious condition is
\begin{equation}\label{cond1}
  \gamma = \beta ,
\end{equation}
because it alone eliminates $\lambda$ in three terms, so that
$\lambda$ will be present only in the terms with $k$ and $m$. We
have to consider several cases of constraints on $k$ and $m$ for
vanishing the remaining dependence on $\lambda$.
\begin{enumerate}
  \item \textbf{Case 1 (generic).}\\
  $\alpha + \beta = 0$ and $\alpha = 0$, so that $\alpha = \beta =\gamma =
  0$ and hence the symmetry generator $X = 0$ in
  (\ref{generator}).There are no symmetries in the generic case.
  \item \textbf{Case 2.}\\
  $k=0$, arbitrary $m$. Then $\beta = \gamma = 0$ and $\alpha$ is
  arbitrary, so it can be set to $1$. The symmetry
  (\ref{generator})becomes
\begin{equation}\label{k=0}
  X = a\frac{\partial}{\partial a}\,.
\end{equation}
  \item \textbf{Case 3.}\\
  $m=0$, arbitrary $k$. Then $\alpha + \beta = 0$ and, since $\gamma =
  \beta$, we can set $\beta = \gamma = 1$ and $\alpha = -1$ in
  (\ref{generator}), so that the symmetry becomes
\begin{equation}\label{m=0}
  X = H\frac{\partial}{\partial H} + F\frac{\partial}{\partial F}
  - a\frac{\partial}{\partial a}\,.
\end{equation}
  \item \textbf{Case 4.}\\
  $k = m = 0$. Then $\beta = \gamma$ is the only condition and we
  have two inequivalent choices:
  \begin{enumerate}
     \item $\beta = \gamma = 0$, $\alpha = 1$ in (\ref{generator})
     with the resulting symmetry
           \begin{equation}\label{X1}
              X_1 = a\frac{\partial}{\partial a}\,,
           \end{equation}
     \item $\beta = \gamma = 1$, $\alpha = 0$ and the symmetry is
           \begin{equation}\label{X2}
              X_2 = H\frac{\partial}{\partial H} + F\frac{\partial}{\partial
              F}\,.
           \end{equation}
  \end{enumerate}
\end{enumerate}
To summarize, we have only one symmetry in cases $2$ and $3$ and
two symmetries in case $4$. One symmetry is not enough for
integrating a system of two first order equations, so in cases $2$
and $3$ we can only find invariant solutions \cite{olv},
particular solutions of our equations. In case $4$ with two
symmetries we can find all solutions of our equations, that is,
integrate our system in quadratures.

\section{Case of flat spacelike sections: \boldmath{$k = 0$}}
\setcounter{equation}{0}
 \label{sec:k=0}

In this case we have only one symmetry (\ref{k=0}) with the basis
of invariants $\{H, F\}$. The equation expressed only in terms of
invariants with solutions of the form $F = F(H)$ can be obtained
by dividing one of equations (\ref{Heq}), (\ref{Feq}) over the
other one, with $k=0$, with the result of the form $dF/dH =
G(H,F)$ which is a first order equation with no more known
symmetries for $m\ne 0$ and hence it cannot be integrated. Though
we cannot find the general solution in case $2$, we still can
search for invariant solutions with respect to the symmetry
(\ref{k=0}), which satisfy the conditions
\begin{equation}\label{eqs2}
  \frac{dH}{da} = 0,\quad \frac{dF}{da} = 0
\end{equation}
and produce a "static" solution $\{H, F\} = constant$. These $H$
and $F$ are roots of the algebraic equations
\begin{gather}
 \hspace*{-33mm}(\nu\omega + 6)H^2 +
 2(\nu + 4)\omega H F + \frac{2\omega}{3}\,[(\nu + 6)\omega -
 3]F^2 \nonumber \\
 \mbox{} + \frac{2\omega}{3}\,(\nu\omega + 3)m^2 = 0,
   \label{algHF}
  \\ \frac{3}{2}\,(\nu + 4)H^2 + 3(2\omega + \nu + 1)H F + [(\nu + 8)\omega - 6]F^2
  + (\nu + 2)\omega m^2 = 0 .\nonumber
\end{gather}
There are four roots of this system of two quadratic equations.
Two of them are real:
\begin{equation}\label{rootsHFreal}
  H = \pm\frac{2m\sqrt{\omega}}{\sqrt{-\omega(\nu^2 + 6\nu) -
  12}}\,,\quad F = \pm\frac{m\nu\sqrt{\omega}}{\sqrt{-\omega(\nu^2 + 6\nu) -
  12}} \; \Longrightarrow\; F = \frac{\nu}{2}\,H ,
\end{equation}
where the expression under the square root in the denominators is
positive for $-6 \le \nu_2 <\nu <\nu_1 \le 0$ with the roots of
the denominator $\nu_{1,\,2} = 3\left(-1 \pm \sqrt{1 -
4/(3\omega)}\right)$ and $\omega > 10^4$. The latter range of
values of $\nu$ includes the interesting values $\nu_{dust} = -3$
and $\nu_{radiation} = -4$. For these $\nu$ and $\omega$, two
other roots of (\ref{algHF}) are imaginary and will not be
considered
\begin{equation}\label{rootsHFimage}
  H = \pm \frac{2im(\omega - 1)\sqrt{\omega}}{\sqrt{6\omega^2 - 17\omega +
  12}}\,, \quad F = \mp \frac{im\sqrt{\omega}}{\sqrt{6\omega^2 - 17\omega +
  12}}\, ,
\end{equation}
where the roots of the denominator are $\omega_1 = 3/2$ and
$\omega_2 = 4/3$, so that $6\omega^2 - 17\omega +
  12 > 0$.

  The equations that define $H$ and $F$
\[ \frac{\dot{a}}{a} = H = constant, \quad \frac{\dot{\phi}}{\phi} = F = \frac{\nu}{2}\,H \]
are integrated to give time dependence of the solution
\begin{equation}\label{sol2}
  a = a_0 e^{Ht},\quad \phi = \phi_0
  e^{{\textstyle\frac{\nu}{2}}\,Ht} .
\end{equation}
The solution (\ref{sol2}) satisfies the initial condition
(\ref{HF1}) if arbitrary constants of integration $\phi_0$ and
$a_0$ are related by the formula
\begin{equation}\label{fi0}
  \phi_0 = \pm\frac{a_0^{\nu/2}}{m}\sqrt{\frac{C}{3}}\,
  \sqrt{\frac{(\nu^2 + 6\nu)\omega + 12}{\nu(\omega - 1) + 1}}\,,
\end{equation}
where $C$ is the coefficient of power law for the density
(\ref{power}).

\section{Massless case: \boldmath{$m = 0$}}
\setcounter{equation}{0}
 \label{sec:m=0}

In this case we also have only one symmetry (\ref{m=0}) with the
basis of invariants $\{\varphi = aF, \psi = aH\}$ (indeed,
$X\varphi = 0,\, X\psi = 0$). Substituting $F = \varphi/a$ and $H
= \psi/a$ into (\ref{Heq}) and (\ref{Feq}) with $m = 0$, we obtain
equations with the new unknowns $\varphi$ and $\psi$:
\begin{eqnarray}
  (2\omega - 3)a\psi\frac{d\psi}{da} &=& [(\nu + 2)\omega +
  3](\psi^2 + k) + 2(\nu + 4)\omega\varphi\psi \nonumber
  \\ & &\mbox{} + \frac{2\omega}{3}\,  [(\nu + 6)\omega - 3]\varphi^2 ,
  \label{phipsi}
  \\ - (2\omega - 3)a\psi\frac{d\varphi}{da} &=& \frac{3}{2}\,(\nu +
  4)(\psi^2 + k) + [4\omega + 3(\nu + 2)]\varphi\psi
   \nonumber
   \\ & &\mbox{} + [(\nu + 8)\omega - 6]\varphi^2 .
\end{eqnarray}
We can obtain only one equation expressed solely in terms of
invariants by dividing one of these equations over another with
the result of the form $d\psi/d\varphi = G(\varphi,\psi)$, but
this equation has no more known symmetries for $k\ne 0$ and hence
it cannot be integrated. Though we cannot find the general
solution in case $3$, again we can search for invariant solutions
with respect to the symmetry (\ref{m=0}) now, which satisfy the
conditions
\begin{equation}\label{eqs3}
  \frac{d\varphi}{da} = 0,\quad \frac{d\psi}{da} = 0 ,
\end{equation}
so that now $\varphi$ and $\psi$ are constants independent of $a$.
With these conditions, equations (\ref{phipsi}) form an algebraic
system:
\begin{eqnarray}
  & &[(\nu + 2)\omega + 3](\psi^2 + k) +
 2(\nu + 4)\omega \varphi \psi + \frac{2\omega}{3}\,[(\nu + 6)\omega -
 3]\varphi^2 = 0 , \nonumber \\
    \label{algphipsi}
  \\ & & 3(\nu + 4)(\psi^2 + k) + 2[4\omega + 3(\nu + 2)]\varphi \psi +
  2[(\nu + 8)\omega - 6]\varphi^2 = 0 ,
  \nonumber
\end{eqnarray}
which again has four roots. Two real roots are
\begin{equation}\label{realroots}
  \varphi = \pm \frac{(\nu + 2)\sqrt{3k}}{\sqrt{-2[(\nu^2 + 8\nu)\omega + 12\omega +
  6]}}\,,\quad \psi = \pm \frac{\sqrt{6k}}{\sqrt{-2[(\nu^2 + 8\nu)\omega + 12\omega +
  6]}}\,,
\end{equation}
that also yields $F = \varphi/a$ and $H = \psi/a$. Formulas
(\ref{realroots}) imply
\begin{equation}\label{relation}
  \varphi = \frac{(\nu + 2)}{2}\,\psi \;\Longrightarrow\; F = \frac{(\nu +
  2)}{2}\, H .
\end{equation}
The expression under the square root in the denominators is
positive for the values of $\nu$ satisfying $-6 \le \nu_2 <\nu
<\nu_1 \le -2$ where the roots of the denominator $\nu_{1,\,2} =
-4 \pm \sqrt{4 - 6/\omega}$ and $\omega > 10^4$. Physically
interesting values $\nu_{dust} = -3$ and $\nu_{radiation} = -4$
again lie in this range, so the roots (\ref{realroots}) are indeed
real. Two other roots are $\varphi = 0 \Rightarrow F = 0
\Rightarrow \phi = \phi_0 = constant$, $\psi = \pm\sqrt{-k}$.
Thus, if $k=0$, $H = F = 0$; if $k = 1$, $H$ is imaginary; if $k =
-1$, $\psi = \pm 1$, $H = \pm 1/a$ and $a = a_0 \pm t$, that is a
trivial solution.

For the first two real roots we have $\dot{a} = \psi = constant$
and so $a = \psi t + a_0$. Then the relation $\dot{\phi}/\phi =
\varphi/a$ yields time dependence of our solution
\begin{equation}\label{fisol}
  \phi = \phi_0 (\psi t + a_0)^{\varphi/\psi} = \phi_0 (\psi t + a_0)^{(\nu +
  2)/2} ,\quad a = \psi t + a_0 ,
\end{equation}
where we have used (\ref{relation}). The solution (\ref{fisol})
satisfies the initial condition (\ref{HF1}) if $\phi_0$ is
expressed in terms of other constants as follows
\begin{eqnarray}
 \phi_0 &=& \pm\frac{2\sqrt{C\omega}}{\sqrt{3(\psi^2 + k)
  + 6\varphi\psi + 2\omega\varphi^2}} \nonumber \\
  &=& \pm \frac{4\sqrt{C\omega}}{\sqrt{(\nu + 2)[(\nu + 2)(2\omega + 3) + 12]\psi^2
  + 12k}} \,,
  \label{fi_0}
\end{eqnarray}
where $C$ is again the coefficient of power law for the density
(\ref{power}).

\section{Massless case with flat spacelike sections: \boldmath{$k = m = 0$}}
\setcounter{equation}{0}
 \label{sec:k=m=0}

In this case we have two commuting symmetries (\ref{X1}) and
(\ref{X2}) which is enough for the complete integration of our
equations in quadratures that will yield their general solution.
We apply first the symmetry generator $X_1 = a\partial/\partial a$
with the invariants $H$ and $F$. A solution expressed solely in
terms of invariants has the form $F = F(H)$. Dividing equation
(\ref{Feq}) over (\ref{Heq}) with $k = m = 0$, we obtain a single
first order equation for $F(H)$:
\begin{equation}\label{F(H)}
  - \frac{dF}{dH} = \frac{[(\nu + 8)\omega - 6]F^2 + 3(2\omega + \nu + 1)HF + (3/2)(\nu + 4)H^2}
  {(2\omega/3)[(\nu + 6)\omega - 3]F^2 + 2(\nu + 4)\omega HF + (\nu\omega +
  6)H^2}\,.
\end{equation}
This equation still admits one symmetry $X_2 =
F\frac{\displaystyle\partial}{\displaystyle\partial F} +
H\frac{\displaystyle\partial}{\displaystyle\partial H}$ with the
invariant $G = F/H$ (and also $a$, not contained explicitly in
(\ref{F(H)})), so that a new independent variable is $H$ and a new
unknown is $G = G(H)$. In the new variables $\{H, G\}$ equation
(\ref{F(H)})) becomes the one with separated variables $H$ and $G$
\begin{gather}
  - \frac{dH}{H} = 2\frac{2\omega[(\nu + 6)\omega - 3]G^2 + 6(\nu + 4)\omega G
  + 3(\nu\omega + 6)}{\{2[(\nu + 6)\omega - 3]G + 3(\nu + 4)\}
  (2\omega G^2 + 6G + 3)} \, dG ,
  \label{G(H)}
\end{gather}
where the cubic polynomial in $G$ in the denominator was
factorized. Integrating both sides of this equation in $H$ and $G$
and getting rid of logarithms, we obtain explicitly the function
$H(G)$
\begin{eqnarray}
  & & H = H_0 \left\langle\{2[(\nu + 6)\omega - 3]G + 3(\nu + 4)\}^\nu (2\omega G^2 + 6G + 3)^3
  \right\rangle^{-b}\times
  \label{H(G)}
  \\ & & \left\{\frac{2\omega G^2 + 6G + 3}{2[(\nu + 6)\omega - 3]G + 3(\nu +
  4)}\right\}^q \exp{\left\{(\nu + 3)l\tan^{-1}{\left[\frac{2\omega G + 3}{\sqrt{3(2\omega -
  3)}}\right]}\right\}}\,,
  \nonumber
\end{eqnarray}
where we have introduced the following constants
\begin{eqnarray}
  & & b = \frac{(\nu + 6)\omega - 3}{(\nu + 6)^2\omega - 6(\nu +
  5)}\,, \quad l = \frac{2\sqrt{3(2\omega - 3)}}{(\nu + 6)^2\omega - 6(\nu +
  5)}\,, \nonumber
  \\ & & q = \frac{3(\nu + 4)}{(\nu + 6)^2\omega - 6(\nu + 5)}\,.
  \label{blq}
\end{eqnarray}
Using (\ref{H(G)}), from the definition of $G$ we also have $F$ as
a function of $G$: $F = GH(G)$.

The dependence $G(a)$ can be determined from (\ref{Heq}) with $F =
GH(G)$ at $k = 0$ and $m = 0$:
\[\frac{da}{a} = \frac{3(2\omega - 3)}{2\omega[(\nu + 6)\omega - 3]G^2
+ 6(\nu + 4)\omega G + 3(\nu\omega + 6)} \times \frac{dH}{H}\,, \]
where $dH/H$ is expressed from (\ref{G(H)}). The resulting
equation is
\begin{equation}\label{da(dG)}
  \frac{da}{a} = \frac{-6(2\omega - 3)dG}{\{2[(\nu + 6)\omega - 3]G + 3(\nu + 4)\}
  (2\omega G^2 + 6G + 3)}\,.
\end{equation}
Integrating both sides of (\ref{da(dG)}) and eliminating
logarithms, we obtain an explicit dependence $a(G)$
\begin{equation}\label{a(G)}
  a = a_0 \left\langle\frac{2\omega G^2 + 6G + 3}{\{2[(\nu + 6)\omega - 3]G + 3(\nu +
  4)\}^2}\right\rangle^b\times \exp{\left\{
  l\tan^{-1}{\left[\frac{2\omega G + 3}{\sqrt{3(2\omega -
  3)}}\right]}\right\}} .
\end{equation}
Formulas (\ref{H(G)}) and (\ref{a(G)}) together with $F = GH(G)$
yield a parametric representation, with a parameter $G$, of the
general solution $H(a), F(a)$ to equations (\ref{Heq}) and
(\ref{Feq}) with $k = m = 0$.

 From $\dot{\phi}/\phi = F = GH$ we have
 \[\frac{d\phi}{\phi} = GHdt = G \frac{da}{a} = \frac{-6(2\omega - 3)GdG}{\{2[(\nu + 6)\omega - 3]G + 3(\nu + 4)\}
  (2\omega G^2 + 6G + 3)}\,. \]
where we have expressed $da/a$ from (\ref{da(dG)}). Integrating
both sides of this equation and eliminating logarithms, we obtain
the explicit dependence $\phi(G)$
\begin{eqnarray}
  & & \phi = \phi_0 \left\{\frac{2[(\nu + 6)\omega - 3]G + 3(\nu + 4)}
  {\sqrt{2\omega G^2 + 6G + 3}}\right\}^q \nonumber \\
  & & \times \exp{\left\{\frac{-(\nu + 6)l}{2}
  \tan^{-1}{\left[\frac{2\omega G + 3}{\sqrt{3(2\omega - 3)}}\right]}\right\}}
  \label{fi(G)}
\end{eqnarray}
and because of the known dependence $a(G)$ from (\ref{a(G)}) we
have determined $\phi(a)$ in a parametric representation. In order
to satisfy the initial condition (\ref{HF1}), the constants in the
formulas (\ref{H(G)}), (\ref{a(G)}), and (\ref{fi(G)}) should be
related as follows
\begin{equation}\label{phi0}
  \phi_0 = \pm 2\sqrt{C\omega}\,\frac{a_0^{\nu/2}}{H_0}\,.
\end{equation}

 To determine time dependence of the solution, we use $\dot{a}/a =
H$ to obtain
 \[t - t_0 = \int\frac{da}{aH} \]
and then express $da/a$ from (\ref{da(dG)}) and $H$ from
(\ref{H(G)}). The integrand simplifies if we introduce the new
parameter $g = (2\omega G + 3)/\sqrt{3(2\omega - 3)}$ with the
resulting integral
\begin{eqnarray}
 \hspace*{-15.5pt}  & & t - t_0 = - \frac{3(2\omega -
  3)}{H_0}\left(\frac{2}{\omega}\right)^{(\nu + 3)b} \int
  \frac{\{[(\nu + 6)\omega - 3]g - \sqrt{3(2\omega - 3)}\}^{2(\nu + 3)b - 1}}
  {(g^2 + 1)^{(\nu + 3)b}} \nonumber \\
 \hspace*{-15.5pt} & & \times \exp{\left[-(\nu + 3)l\tan^{-1}{g}\right]}\,dg,
  \label{t(G)}
\end{eqnarray}
that gives us explicitly functions $t(g)$ and hence $t(G)$ and
implicitly the time dependence $G(t)$. Then from the functions
$a(G)$ and $\phi(G)$, determined by (\ref{a(G)}) and (\ref{fi(G)})
respectively, we know the time dependencies $a(t)$ and $\phi(t)$
in a parametric representation with the parameter $g$ in terms of
the quadrature (\ref{t(G)}).

 For dust, $\nu = \nu_{dust} = -3$ and the integral (\ref{t(G)})
simplifies substantially with the result
\begin{equation}\label{dust}
  t - t_0 = -\frac{(2\omega - 3)}{(\omega - 1)H_0}\ln{\left[3(\omega - 1)g
  - \sqrt{3(2\omega - 3)}\right]} ,
\end{equation}
that can be inverted easily to yield $g(t)$ and so an explicit
expression for $G(t)$:
\begin{equation}\label{G(t)}
  G = \frac{1}{2\omega(\omega - 1)}\left\{\sqrt{\frac{2\omega - 3}{3}}
  \exp{\left[-\frac{(\omega - 1)}{(2\omega - 3)}\,H_0(t - t_0)\right]} -
  \omega\right\}.
\end{equation}
 For radiation, $\nu = \nu_{radiation} = -4$ and the simplification is
not enough for evaluating the integral in (\ref{t(G)}).

\section{Conclusion}
\setcounter{equation}{0}
 \label{conclude}

    We have shown that, in the model of cosmology we are
 considering, the equation of state does not affect the expansion
 law of the universe. Thus, for the $m=0$ case the expansion of the
 universe, following the invariant solution (\ref{fisol}), is always linear. What is interesting
 is that according to (\ref{realroots}) this solution is nontrivial only for
 k=1 i.e. a closed universe. It is also interesting to note that the equation of state
 which determines the coefficient $\nu$ affects the rate of
 expansion. This case may be relevant for the early universe where
 the curvature term cannot be neglected. (\ref{fisol}) shows that with $a_0=0$, for
 $t<<m^{-1}$
 \begin{equation}\label{eq}
   \frac{\dot{\phi}}{\phi} = \frac{\nu + 2}{2}\,\frac{1}{t} >>
   \frac{\nu + 2}{2}\, m ,
\end{equation}
 so that $m|\phi|<<|\dot{\phi}|$ and for early times the mass term
 can indeed be neglected.
    After the universe expands to a large size, the curvature term
    can be neglected and in this case the k=0 solution becomes
    important. (\ref{rootsHFreal}), (\ref{sol2}) show that the universe expands
    exponentially for all equations of state. The
    equation of state is important just for the value of the
   Hubble constant. This behavior we identify with the dark
   energy or cosmological constant which is observed today.

\section*{Acknowledgements}

The research of the authors is partly supported by the research
grant from Bogazici University Scientific Research Fund, research
project No. 07B301.


\begin{thebibliography}{99}
\bibitem{roos}
 Roos M, {\it Expansion of the Universe - Standard Big Bang
Model}, 2008, {\it Preprint} astro-ph/0802.2005
\bibitem{PenWils}
 Penzias A A and Wilson R W, 1965 {\it Astrophys.
J.} {\bf 142} 419
\bibitem{Fried}
 Friedman A, 1922 {\it  Z. Phys.} {\bf 10} 377 (in German) (English translation in: Friedman
A, 1999 {\it Gen. Rel. Grav.} {\bf 31} 1991)
\bibitem{Lem}
  Lema\^{\i{}}tre G, 1927 {\it Ann. Sci. Soc. Brussels} {\bf 47A}
41 (in French) (English translation in: Lema\^{\i{}}tre G, 1931
{\it Monthly Notices Royal Astronom. Soc.} {\bf 91 } 483)
\bibitem{JTBD}
 Jordan P, 1947 {\it Ann. Phys. (Leipzig)} {\bf 1} 219; Thirry Y
R, 1948 {\it C. R. Acad. Sci. (Paris)} {\bf 226} 216; Brans C and
Dicke C H, 1961 {\it Phys. Rev.} {\bf 124} 925
\bibitem{DT}
 Dereli T and Tucker R W, 1983 {\it Phys. Lett.} {\bf 125B} 133
\bibitem{olv}
Olver P, 1986 {\it Applications of Lie Groups to Differential
Equations} (New York: Springer-Verlag)
\end{thebibliography}
\end{document}